# Dislocation Pinning in Helium-Implanted Tungsten: A Molecular Dynamics Study

*Suchandrima Das[*a], Andrea Sand [b], Felix Hofmann[**c]*

[a] *Department of Mechanical Engineering, University of Bristol, Bristol, UK*

[b] *Department of Applied Physics, Aalto University, Finland*

[c] *Department of Engineering Science, University of Oxford, Oxford, UK*

[*]*suchandrima.das@bristol.ac.uk*

[**]*felix.hofmann@eng.ox.ac.uk*

## Abstract

The interaction of edge dislocation with helium-implantation-induced defects in tungsten is investigated using molecular dynamics. Following prior investigations, we consider defects with two helium ions in a vacancy with a self-interstitial bound to it ($He_2$V-SIA). Our observations suggest 3-10 $He_2$V-SIA cluster together, with their pinning strength on glide dislocations increasing with size. For all cluster sizes, the dislocation bows around the cluster, until it gets unpinned, carrying the SIAs with it and leaving behind a helium-vacancy complex and newly created vacancies in its wake. The remnant helium-vacancy complex has little pinning effect, highlighting the "defect-clearing" process. A total solute hardening force for a distribution of clusters of different sizes, induced by 3000 appm of helium, is found to be ~700 MPa. This is in good agreement with the corresponding value of 750 MPa estimated in a previously developed crystal plasticity model simulating the deformation behaviour of the helium-implanted tungsten.

## Main Text

Tungsten is a front-runner candidate for armour components in future fusion reactors [1–3]. Helium-ion-implanted tungsten is widely used to study changes in material properties induced by in-service irradiation with neutrons and helium [4–16]. Tungsten implanted with helium at room temperature allows investigation of helium-induced damage with little damage evolution, as vacancies in tungsten are effectively immobile at room temperature [17,18]. Investigation through various experimental techniques such as thermal desorption spectroscopy [19], nuclear reaction analysis [20] and positron annihilation spectroscopy [21] has shown that in this regime, helium binds strongly to vacancies and vacancy clusters in tungsten. Monochromatic X-ray diffraction (XRD) measurements (strain resolution $\sim 10^{-4}$) combined with density functional theory (DFT) calculations predicts such defects to be constituted of di-helium vacancy complexes bound to a self-interstitial (SIA) ($He_2V$-SIA) [5,13,22,23]. Unfortunately transmission electron microscope (TEM) is not sufficiently sensitive to pick up these few atom large defects [6,24].

Though small, helium-induced-defects can significantly modify material structure and properties. Tungsten implanted with 3000 appm of helium (W-3000He) exhibited large lattice strains ($\sim 10^{-3}$) [25], as well as a substantial reduction in thermal diffusivity [26]. Nano-indentation on <001>-oriented W-3000He showed a ~70% increase in hardness and slip-step formation around indents [9]. A localisation of deformation around and beneath indents was revealed by depth-resolved micro-beam XRD and high-resolution electron backscatter diffraction [14]. TEM investigation revealed slip channels beneath indents, similar to other irradiated materials [14,27,28].

Based on these observations, we previously developed a crystal plasticity finite element model (CPFE) to simulate deformation behaviour of W-3000He [14]. The formulation considered hardening from $He_2V$-SIA obstacles, followed by defect removal and strain-softening [9,14,29]. Validation of this formulation against nano-indentation experiments showed good quantitative agreement. These simulations also correctly reproduced the orientation-dependence of nano-indentation response, and were further extended to self-ion-implanted tungsten [29,30].

The CPFE formulation calculated the pinning strength of the He$_2$V-SIA defects by considering that gliding dislocations remove the helium from the He$_2$V-SIA complex, helping the recombination of the vacancy and SIA and creating channels with reduced defect density [13,14,31]. Quantitative agreement of the experimental observations with the numerical predictions suggests that hardening and subsequent deformation-induced softening are the right mechanisms to consider. However, how exactly these operate at the nano-scale is unclear.

Here, we address this question using molecular dynamics (MD). MD has been extensively used to study interaction of point defects, helium, helium-vacancy complexes, loops, helium bubbles etc. with dislocations in metals [23,32–37]. The focus of these studies has been on the binding energy of defects to dislocations or on the migration of defects in the vicinity of dislocation[32]. The interaction between pinning interstitial loops and gliding dislocations has also been considered to evaluate irradiation hardening [34–36]. Here we use MD to examine the interaction between gliding dislocations and He$_2$V-SIA defects at 300 K [5,13,22,23]. Plasticity in pure tungsten is likely controlled by the screw dislocations with their relatively lower mobility than edge dislocations [38]. However, ab-initio calculations have shown that in tungsten, the interaction energy of helium with the core of the edge dislocation is almost double that with screw dislocation [39]. Hence, here we concentrate on investigating the interaction between He$_2$V-SIA defects and edge dislocation, which have the highest mobility.

The embedded atom potential developed by [40] was used to simulate a tungsten crystal (with lattice constant 3.15 Å, further details in Appendix A) containing helium defects. It is necessary to establish a viable defect configuration whose interaction with dislocations and the associated pinning strength can be examined. Thus, we investigate the stability and possible configuration of the He$_2$V-SIA defect by relaxing it in an otherwise perfect tungsten cubic cell (edge dimension 62 Å) (Figure 1 (a)). As a detailed examination of the defect energy landscape is beyond the scope of this study, this is done here by investigating the binding energy (BE) between the SIA and the $He_2V$ complex.

The $He_2V$ defect is generated by removing one tungsten atom to create a vacancy, followed by introduction of two helium ions in the vacancy (if the vacancy is at co-ordinates [0,0,0], the helium ions are introduced at [0.5,0,0] and [-0.5,0,0], where dimensions are in Å). Previous work predicting the $He_2V$-SIA configuration did not specify the position of the SIA relative to the $He_2V$ [5]. Thus, BE is calculated for different positions of the SIA relative to the $He_2V$ to identify a viable configuration (details in Appendix B). The highest BE of 0.04 eV is found for the SIA positioned 3.95 Å away from the vacancy in the X and Z directions (XYZ directions shown in Figure 1 (a)) and 0.78 Å away in the Y direction (configuration shown in Figure 1 (b)-(c)). With the low BE, this configuration is unstable at 300 K.

Boisse et al. has shown that the BE of SIAs with helium-vacancy (He-V) complexes increases with increasing helium in the He-V complex [13]. Becquart et al., suggests that in tungsten, the stability of the He-V complex itself, increases with increasing associated vacancies [41]. These observations and our observation of unstable $He_2V$-SIA configuration motivated the search for a defect configuration made up of clusters of $He_2V$-SIA.

To create a cluster of size N, N neighbouring atoms were removed to create N vacancies (neighbouring atoms with co-ordinates within 8 Å in the X, Y and Z directions were chosen (Figure 1 (a) shows co-ordinates). Relative to each created vacancy, two helium ions and a SIA were introduced. During potential energy minimisation, neighbouring $He_2V$-SIA clustered, causing helium ions to relax into fewer vacancies than originally introduced and allowing recombination of the isolated vacancies with nearby SIAs (Figure 1 (d)-(i)). For example, for N = 3, of the three Frenkel Pairs introduced, one was annihilated leaving two SIAs and two vacancies to accommodate the six helium ions. 1<=N<=10 was explored as the induced defects are all below the TEM visibility limit [6]. The results show that BE (details in Appendix B) increases dramatically for N>2 and stabilises to ~3 eV thereafter (Figure 1 (j)).

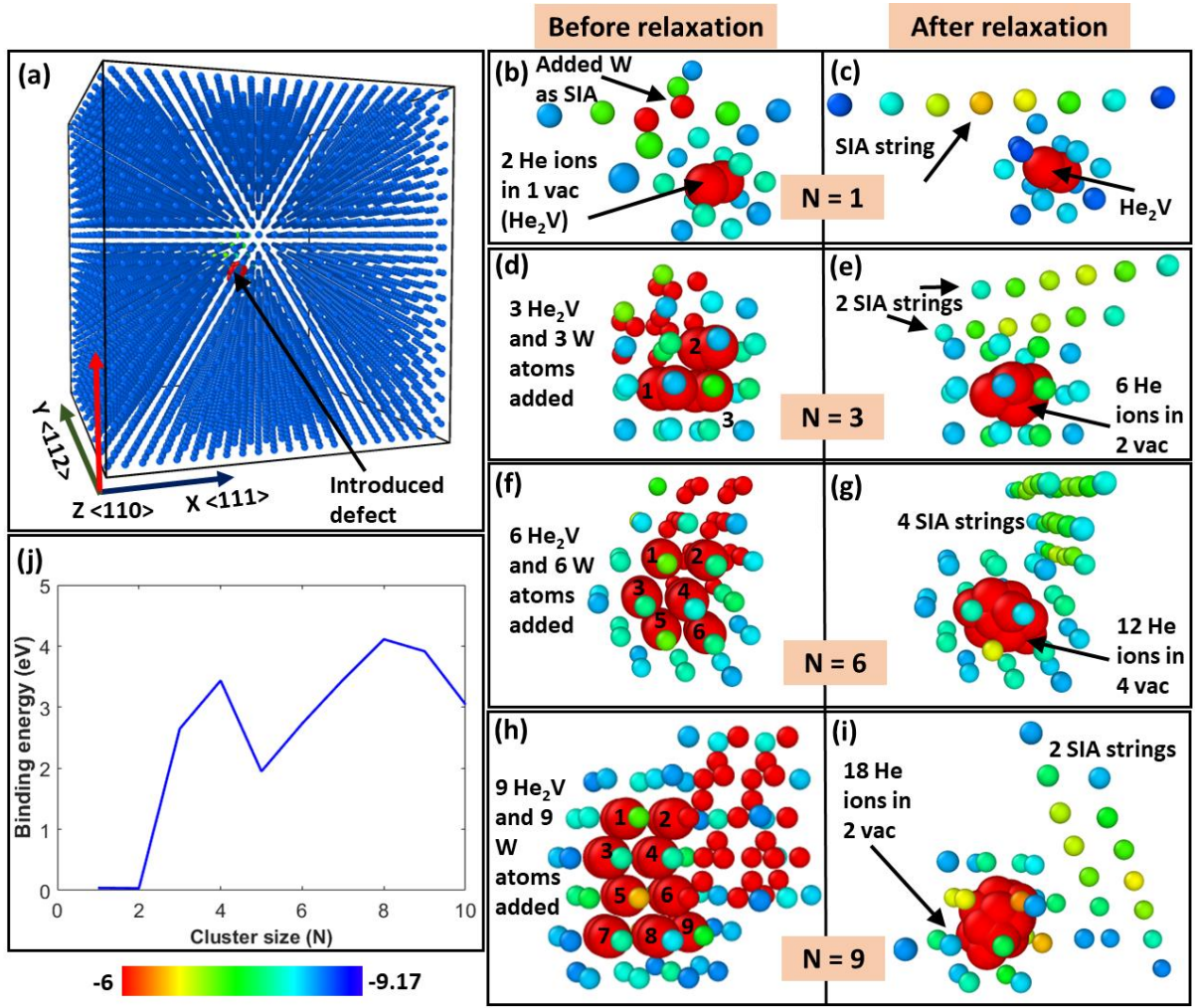

Figure 1 – (a) Cubic simulation cell (edge dimension 62 Å) of tungsten atoms with X, Y, Z co-ordinates superimposed used to examine stability of defects. (b) – (i) Examples of formation of $(He_2V\text{-}SIA)_N$ clusters with N varying from 1 to 9. (j) Variation of binding energy, SIA to $He_2V$, plotted as a function of cluster size N. In subplots (a) – (i), the radius of tungsten atoms has been reduced to 0.6 Å for ease of visualisation of the helium atoms (shown larger in the figures with radius 1.22 Å) and the induced defects. Subplots (a) – (i) are coloured according to potential energy (in units of eV) of the individual atoms, with range as indicated in the colour bar below subplot (j).

Figure 1 (j) suggests that $He_2V$-SIA clusters of N>=3 are stable. This is confirmed by allowing the different cluster sizes to equilibrate for 3 ns at 300 K. The configuration is stable for N>=3. For N<3, after a few ps the SIAs detach from the immobile $He_2V$ cluster and become mobile. This confirms the presence of clusters of $He_2V$-SIA defects with N>=3 in W-3000He.

Next, we consider the interaction of $He_2V$-SIA clusters with a dislocation. A simulation cell with an edge dislocation (with Burgers' vector ½[111] (X-axis), line direction $[11\bar{2}]$ (Y-axis) and on slip plane

[1$\bar{1}$0] (normal to Z-axis)) introduced in the middle is considered (Figure 2 (a)). Periodic boundary conditions are applied along the X and Y directions to simulate an infinite array of parallel dislocations and improve the accuracy of defect-dislocation interaction by minimising strain on the dislocation core [42] (details in Appendix C). Cell dimensions (Figure 2 (a)) are chosen to minimise dislocation-image interactions and to obtain a dislocation density of ~$10^{15}$ m$^{-2}$, similar to that encountered generally in plastically deformed metals [43]. The cell contains 944463 atoms. Two blocks of atoms, in the top (Block A) and bottom (Block B) of the cell (Figure 2 (a)) are kept immobile. The defect cluster is introduced in the path of the dislocation, 150 Å away from it, along the X-axis (Figure 2 (a)). Eight simulations were carried out, for 3<=N<=10. Following potential energy minimisation, the cell is equilibrated for 250 ps at 300 K. Subsequently, Block B is kept fixed while Block A is displaced to give a constant shear strain rate of $10^{-6}$ ps$^{-1}$. This shear strain rate minimises the rate dependence of pinning strength (details in Appendix D). The shear stress required to achieve the pre-defined shear strain is continuously recorded until the dislocation passes the defect and reaches the edge of the cell. The shear stress is multiplied with the surface area of the cell's XY face ($1.47 \times 10^{-15}$ m$^2$) (Figure 2 (a)) to obtain the pinning force.

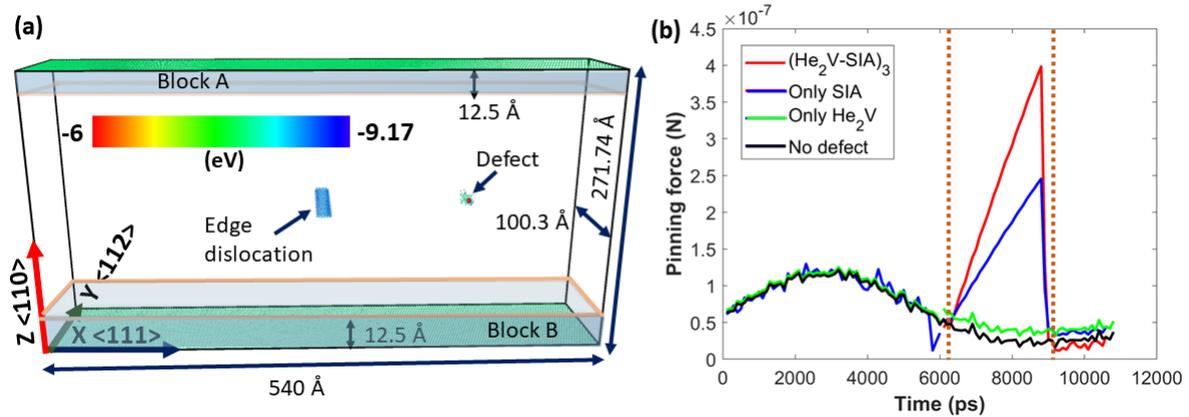

Figure 2 – (a) Simulation cell with edge dislocation (½[111] Burgers' vector, [1$\bar{1}$0] slip plane and [11$\bar{2}$] line direction) at the cell centre and a He$_2$V-SIA defect cluster introduced in its path. The atoms in the cell are coloured according to their potential energy (in units of eV) with range as indicated in the colour-bar. Atoms with potential energy < -8.7 eV are excluded for ease of visualisation and tungsten atoms are shown with a reduced radius of 0.6 Å. Blocks of atoms above and below the dotted enclosure at the top (block A) and bottom (block B) of cell respectively are kept immobile. (b) Plot of pinning force (averaged over every 100 ps) recorded as the dislocation moves through the cell when block B is kept fixed and block A is sheared at a constant rate of 10$^{-6}$ ps$^{-1}$. Plots are shown for a cell with defect cluster of size three (configuration in Figure 1(d)-(e)); a cell with just 2 SIAs (same as number of SIAs for relaxed defect cluster of size three); a cell with three He$_2$V introduced and relaxed in the same manner as for obtaining a He$_2$V-SIA cluster of size three; and a perfect, defect-free cell.

Figure 2 (b) illustrates the pinning effect by considering a cell with and without a defect cluster of size three. It is noted here that observations made from Figure 2 (b) and subsequently from Figure 3 apply to all 3<=N<=10. Figure 2 (b) shows the measured pinning force averaged over every 100 ps (details in Appendix E). In contrast to a rather flat curve recorded for dislocation glide in a defect-free cell, a clear peak in the curve is noticed in presence of a (He$_2$V-SIA)$_3$ defect, indicating the pinning and unpinning phase. Pinning strength is calculated by subtracting the pinning force for the defect-free case (averaged over the zone demarcated by the orange dotted line in Figure 2 (b)) from the maximum pinning force recorded with the defect present (Table 1). The pinning strength is also calculated by considering the pinning force curves (such as in Figure 2 (b)) averaged over every 20 ps, 50 ps, 200 ps and 250 ps. The ±1 standard deviation across the five values (for the five averaging periods, 20, 50, 100, 150 and 250 ps) is the uncertainty in the calculated pinning strength (Figure 3 (k) and Table 1).

Two further simulations are run with only 2 SIAs (as expected for relaxed (He$_2$V-SIA)$_3$) and with only ((He$_2$V)$_3$) defect (introduced at the same co-ordinates as the (He$_2$V-SIA)$_3$) in the dislocation path respectively. For simulation with only SIAs, the cell was equilibrated for 2 ps at 300 K (instead of 250 ps as for other cases) to prevent the mobile SIAs from moving away. It is seen that a (He$_2$V)$_3$ defect, on its own, has almost no pinning effect, while the pinning strength of the SIAs alone is ~55% that of the (He$_2$V-SIA)$_3$ (Figure 2 (b)). This confirms that SIAs in stable bonding with a He$_2$V cluster result in maximum dislocation pinning effect. This agrees well with nano-indentation measurements, where the increase in hardness in W-3000He (~0.2 dpa) is almost double that of 1 dpa self-ion implanted tungsten (exposed at room temperature and with predominantly SIA loops as defects) [9,30].

Interestingly, for all cluster sizes (3<=N<=10), the dislocation bows around the cluster, until it gets unpinned, carrying the SIAs with it and leaving behind a helium-vacancy complex and newly created vacancies in its wake (Figure 3 (a) – (i)). The dislocation acting as a sink for the SIAs, experiences climb in the process. Since a helium-vacancy complex, without SIA, has no pinning effect (Figure 2 (b)), by carrying away the SIAs, the dislocation removes the pinning characteristic of the defects, creating an effectively "defect-free" channel. This may explain the observation of defect-free channels beneath indents in W-3000He [14]. The observations also suggest the inability of vacancies to pin gliding dislocations. This is in agreement with a past crystal-plasticity study on self-ion implanted tungsten, which captured the experimental observations well by considering only dislocation loops as obstacles and ignoring vacancies [30].

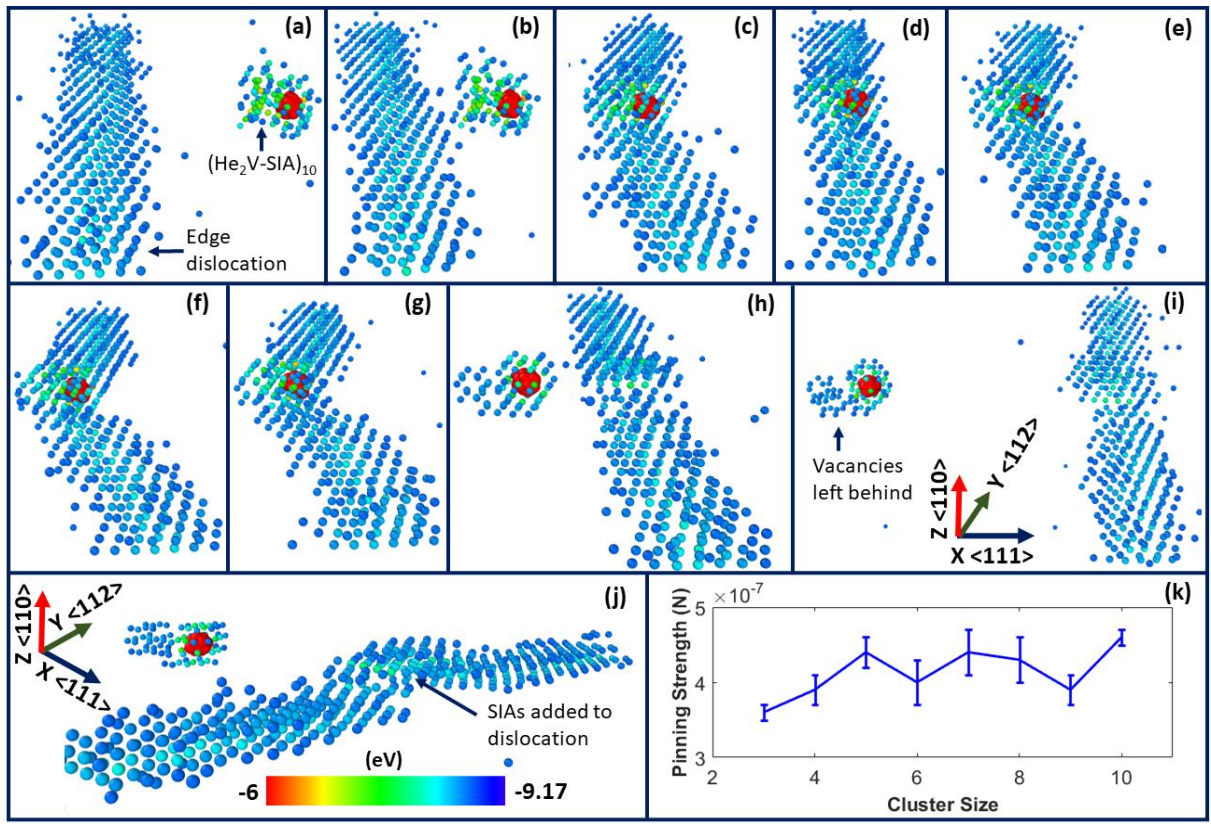

Figure 3 – (a) – (i) Subplots showing defect-dislocation interaction as the dislocation encounters a $(He_2V-SIA)_{10}$ cluster, gets pinned, bows around cluster, unpins and moves past. The atoms are coloured according to their potential energy (in units of eV) with range as indicated by colour-bar enclosed in subplot (j). The XYZ co-ordinates in subplot (i) apply to all subplots from (a) – (i). (j) A rotated view of the dislocation and $(He_2V-SIA)_{10}$ cluster after the dislocation has passed through the defect, demonstrating the SIAs added to the dislocation and helium-vacancy complex and newly created vacancies nearby, left behind in the wake of the dislocation. For ease of visualisation, tungsten atoms in subplots (a) – (j) are plotted with a radius of 0.6 Å, and helium atoms are shown with a radius of 1.22 Å. The same defect-dislocation interaction mechanism was observed for all investigated cluster sizes (N from 3 to 10) (k) Plot of the pinning strength as a function of cluster size. The error bars for each point represent ±1 standard deviation across the values estimated from the measured pinning force data averaged over every 20, 50, 100, 150 and 250 ps respectively.

The pinning strength increases slightly with cluster size (Table 1), with a maximum increase of ~25% (Figure 3 (k)). These values are then used to estimate the total solute hardening force experienced by a gliding dislocation when moving through a concentration of such clusters, induced by 3000 appm of helium, for comparison with the corresponding estimate of 750 MPa from the previous CPFE simulations [14].

A first estimate is made by assuming a uniform defect distribution (similar assumption made in CPFE). To account for the possibility of existence of clusters of varying sizes, the total solute hardening force is calculated separately for a uniform distribution of clusters of each size ($F_{tot}$ in Table 1) followed by taking the average over all cluster sizes. Contrary to pinning strength, $F_{tot}$ is seen to reduce with increasing cluster size, because of the reduction in cluster number density as N increases. The average of $F_{tot}$ for different sizes ($F_{mean}$), is found to be ~700 MPa (Table 1) which is in good agreement with the estimate of 750 MPa from CPFE.

| Cluster size (N) | Pinning strength (F in μN) | Pinning stress $F_1$ = F/ (surface area of XY face of simulation box) (in MPa) | Total no. of helium ions. A = 2 × N | Concentration of clusters for 3000 appm implanted helium B = (3000/A) appm | Spacing between defects considering uniform distribution (C in Å) | No. of defect clusters at any point along slip direction. D = (100/C) | Total solute hardening force. $F_{tot}$ = D × $F_1$ (MPa) |
|---|---|---|---|---|---|---|---|
| 3 | 0.36±0.01 | 250±10 | 6 | 500 | 30.87 | 3 | 750±30 |
| 4 | 0.39±0.02 | 270±10 | 8 | 375 | 33.98 | 3 | 810±30 |
| 5 | 0.44±0.02 | 300±10 | 10 | 300 | 36.61 | 3 | 900±30 |
| 6 | 0.4±0.03 | 270±20 | 12 | 250 | 38.9 | 3 | 810±60 |
| 7 | 0.44±0.03 | 300±20 | 14 | 214.59 | 40.95 | 2 | 600±40 |
| 8 | 0.43±0.03 | 290±20 | 16 | 187.5 | 42.82 | 2 | 580±40 |
| 9 | 0.39±0.02 | 270±10 | 18 | 166.67 | 44.53 | 2 | 540±20 |
| 10 | 0.46±0.01 | 310±10 | 20 | 150 | 46.12 | 2 | 620±20 |
| | | | | | | | **Mean ($F_{mean}$) = 700±124** |

Table 1 – Variation of pinning strength and total solute hardening force with cluster size of $He_2V$-SIA defects. The uncertainty for the calculated pinning strength ($F$) for each cluster size, is ±1 standard deviation across the four values estimated from the measured pinning force data averaged over every 20, 50, 100, 200, 250 ps respectively. The uncertainty for $F_{mean}$ is calculated as ±1 standard deviation across the $F_{tot}$ values for all the cluster sizes considered. The surface area of the XY face of the simulation box (Figure 2 (a)), used in the computation of $F_1$ is $1.47 \times 10^{-15}$ m². Details of calculation of parameters C and D in Appendix F.

To examine the credibility of $F_{tot}$ for varying N (and therefore $F_{mean}$) estimated from uniform defect distribution, $F_{tot}$ is also estimated for N = 3 (500 clusters) and N = 10 (150 clusters), with randomly distributed clusters (Table 1). For each cluster size five simulations were performed. A distance of 540 Å traversed by the dislocation was recorded, such that it did not pass over the same region in the cell twice (Figure 4 (a)-(b)). Block B is kept fixed while Block A is sheared to give a constant strain rate of $10^{-6}$ $ps^{-1}$ (Figure 2 (a)). As expected, the dislocation carries away the SIAs of all the clusters it encounters, leaving behind an SIA-free path in its wake, populated with helium-vacancy complexes and newly created vacancies (Figure 4 (c)-(d)).

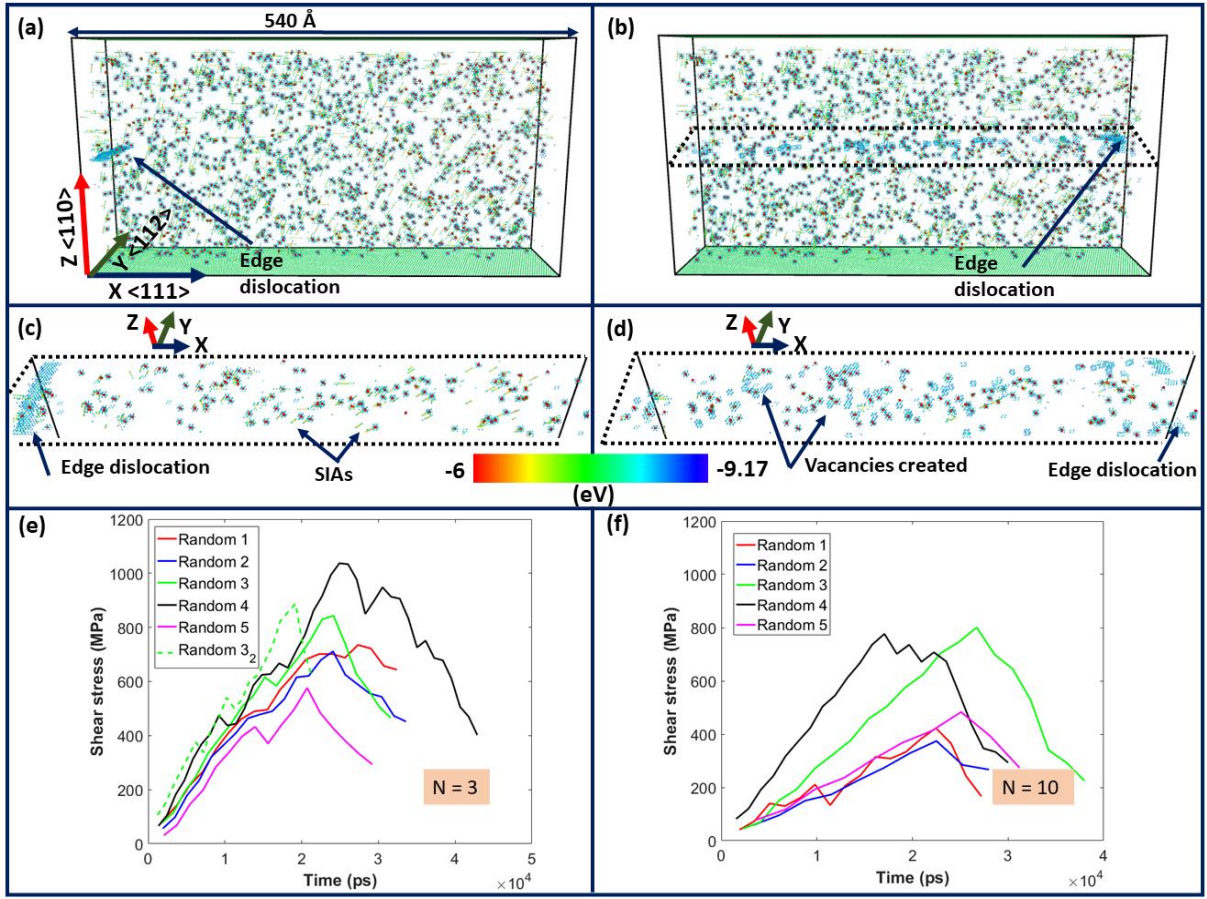

Figure 4 – Simulation cell with random distribution of defect clusters of size N= 3 (a) at the start of the simulation, with the edge dislocation introduced at the edge of the cell (b) at the end of the simulation, when the dislocation has reached the other edge of the cell. XYZ co-ordinates superimposed on subplot (a) also apply to (b). Isolated view of the XY plane containing the gliding dislocation (c) at the beginning of the simulation showing defect clusters with SIAs attached and (d) at the end of the simulation, when the dislocation has passed through the defects, carrying the SIAs with it and leaving behind an SIA-free channel containing the helium-vacancy complexes and newly created vacancies nearby shown in Figure 3. The XYZ axis superimposed on (c) and (d) have the same directions as in (a). Subplots (a) – (d) show atoms coloured according to their potential energy with range indicated by the colour-bar at the bottom of plots (c) and (d). For ease of visualisation, in plots (a)-(d), tungsten atoms are plotted with a radius of 0.6 Å, and helium atoms are shown with a radius of 1.22 Å and atoms with potential energy < -8.7 eV are excluded. Shear stress required for an edge dislocation to glide through the simulation cell for five different random distribution of $He_2V$-SIA defect clusters (a) for cluster size 3 (b) for cluster size 10. The "Random $3_2$" curve in subplot (e), represents the case for when the simulation corresponding to "Random 3" was repeated with the dislocation introduced in the middle of the cell (similar to in Figure 2(a)) i.e. traversed half the length of the cell. Curves in subplots (e) – (f) are averaged over every 100 ps.

Figure 4 (e)-(f) shows the shear stress required for the dislocation to glide through each of the five random distributions for N=3 and N=10. One of the random distributions for N = 3 (Random 3) was repeated with the dislocation traversing half the length of the cell (from middle to the edge, Random $3_2$) (Figure 4 (e)). The peak stress for "Random 3" and "Random $3_2$" being within 5% of each other, confirms that the peak stress is independent of the path-length traversed by the dislocation. Between the random cases, the peak stress reached varies; between ~374-800 MPa with an average of 571.59 MPa for N=10 and between ~577-1037 MPa, with an average of 780.78 MPa for N=3.

The average stress values for N=3 and N=10 are within ~10 % of the corresponding $F_{tot}$ computed from uniform defect distribution (Table 1). Though a quantitative comparison is not justified, by 5 random distributions examined for each N, this qualitative agreement lends credence to the cluster size dependent $F_{tot}$ estimated in Table 1.

The present simulations confirm the pinning effect associated with $He_2$V-SIA defects predicted in W-3000He [9]. Observations of poor stability for N<3 and the indistinct nature of defects in TEM suggest the presence of small clusters of $He_2$V-SIA instead of discrete $He_2$V-SIA defects. The large pinning strength of ~0.4 µN even for N= 3, explains the ~75% increase in hardness observed through nano-indentation of W-3000He [9]. SIAs being carried away by the gliding dislocation and the lack of pinning strength of the remaining $He_2$V clusters explains the mechanism for the formation of slip channels observed in experiments.

Though the defect removal mechanism assumed in the CPFE study in [14] is different from the findings here, there is good agreement in $F_{mean}$. Thus, lack of physical basis may cause CPFE's inability to predict channel formation or orientation-dependent evolution of dislocation structures [44]. But it can still produce results in quantitative agreement with experimental observations of hardness, strain etc [14]. As such, our observations re-affirm the need to use multi-scale tools to fully reproduce radiation damage effects. The future outlook would be to thus implement a CPFE formulation by pooling in the physical understanding of evolving dislocation structures and microstructure from MD and discrete dislocation (DD) plasticity simulations [45,46].

## Acknowledgements

This work was in part supported by the European Research Council through grant 714697 and in part by Engineering and Physical Sciences Research Council fellowship Reference number EP/N007239/1. We are grateful to Dr. Edmund Tarleton for supporting this work.

# Appendix A

The potential considers the lattice constant of tungsten to be 3.14 Å at 0 K. Since the simulations are intended to be representative of phenomena at 300 K, the lattice constant relevant to this potential at 300 K was determined. For this purpose, a cubic cell, made of tungsten atoms without any defects, with edge dimension of 30 nm, was considered. The cell was allowed to achieve potential energy minimisation, following which it was allowed to equilibrate at 300 K for 600 ps. The lattice constant was recorded throughout the simulation (Figure 5) and the average taken over the last 300 ps (3.15 Å) was used in all subsequent simulations.

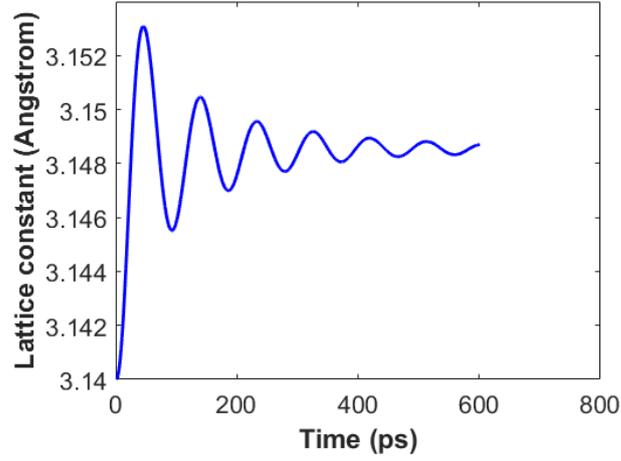

Figure 5 – Plot of change in lattice constant of perfect tungsten cell as it equilibrates at 300 K.

## Appendix B

The study predicting the He$_2$V-SIA configuration does not clarify the position of the SIA relative to the He$_2$V [5]. Thus, the BE for different positions of the SIA relative to the He$_2$V is examined to identify a viable configuration. For example, for a vacancy created at co-ordinates [0,0,0], the SIA is positioned at [X,Y,Z], with X and Z ranging 3.15 to 13.15 Å and -3.15 to -13.15 Å and Y ranging from 0.7 to 13.15 Å and from -0.7 to -13.15 Å (with incremental step-size of 0.2 Å). The lower limits for the ranges of X, Y and Z were chosen based on the minimum distance required to prevent the SIA from annihilating the vacancy upon relaxation. The BE ($E^{BE}_{He_2V-SIA}$) is calculated for each considered position of the SIA relative to the He$_2$V as per Eq. 1

$$\begin{aligned} E^{BE}_{He_2V-SIA} &= E^{form}_{He_2V} + E^{form}_{SIA} - E^{form}_{He_2V-SIA} \\ &= E^{tot}_{He_2V} + E^{tot}_{SIA} - E^{tot}_{He_2V-SIA} - E^{tot}_{pure} \end{aligned} \quad (1)$$

where, $E_{pure}^{tot}$, $E_{He_2V}^{tot}$, $E_{SIA}^{tot}$ and $E_{He_2V-SIA}^{tot}$ are the total potential energy of a perfect tungsten cell, a tungsten cell with He₂V defect, a cell with SIA and a cell with He₂V-SIA respectively and $E_{He_2V}^{form}$, $E_{SIA}^{form}$ and $E_{He_2V-SIA}^{form}$ are the formation energies of a He₂V, SIA and He₂V-SIA defects respectively.

Eq. (1) is derived based on the approach outlined below.

The formation energy of a He₂V ($E_{He_2V}^{form}$) is calculated as

$$E_{He_2V}^{form} = E_{He_2V}^{tot} - E_{pure}^{tot} + E_W - 2E_{He} \qquad (2)$$

where, $E_W$ is the energy of a single tungsten atom and accounts for the single vacancy (i.e. tungsten atom removed), $E_{He}$ the energy of an isolated helium ion and $E_{pure}^{tot}$ and $E_{He_2V}^{tot}$ are the total potential energy of a perfect tungsten cell and a tungsten cell with He₂V defect respectively.

The formation energy of a SIA ($E_{SIA}^{form}$) is calculated as

$$E_{SIA}^{form} = E_{SIA}^{tot} - E_{pure}^{tot} - E_W \qquad (3)$$

where, $E_W$ is the energy of a single tungsten atom and accounts for the introduced SIA and $E_{pure}^{tot}$ and $E_{SIA}^{tot}$ are the total potential energy of a perfect tungsten cell and a cell with SIA respectively.

The formation energy of a He₂V-SIA ($E_{He_2V-SIA}^{form}$) defect is calculated as

$$E_{He_2V-SIA}^{form} = E_{He_2V-SIA}^{tot} - E_{pure}^{tot} + E_W - E_W - 2E_{He} \qquad (4)$$

where, $E_W$ is the energy of a single tungsten atom and accounts for the single vacancy (i.e. tungsten atom removed) and the introduced SIA, $E_{He}$ the energy of an isolated helium ion and $E_{pure}^{tot}$ and $E_{He_2V-SIA}^{tot}$ are the total potential energy of a perfect tungsten cell and a cell with He₂V-SIA respectively.

As per Eq. 1, the binding energy of the He₂V-SIA is calculated as $E^{BE}_{He_2V-SIA} = E^{form}_{He_2V} + E^{form}_{SIA} - E^{form}_{He_2V-SIA}$. Substituting the expressions for $E^{form}_{He_2V}$, $E^{form}_{SIA}$ and $E^{form}_{He_2V-SIA}$ in this equation and simplifying gives

$$E^{BE}_{He_2V-SIA} = E^{tot}_{He_2V} + E^{tot}_{SIA} - E^{tot}_{He_2V-SIA} - E^{tot}_{pure} \qquad (5)$$

For a cluster of He₂V-SIA of size N, the BE was calculated as

$$E^{BE}_{(He_2V-SIA)_N} = E^{tot}_{(He_2V)_N} + E^{tot}_{(SIA)_N} - E^{tot}_{(He_2V-SIA)_N} - E^{tot}_{pure} \qquad (6)$$

where, $E^{tot}_{pure}$, $E^{tot}_{(He_2V)_N}$, $E^{tot}_{(SIA)_N}$ and $E^{tot}_{(He_2V-SIA)_N}$ are the total potential energy of a perfect tungsten cell, a tungsten cell with N He₂V defects, a cell with N SIAs and a cell with N He₂V-SIA defects respectively.

Figure 1 (j) shows the BE calculated for 1<=N<=10 cluster sizes. Small variations in the BE can be attributed to the variance of helium-vacancy complex across cluster sizes. For example, while both N = 3 and N = 9 clusters have 2 SIAs, the latter has slightly higher BE, owing to the larger strain field of its larger helium-vacancy complex (Figure 1 (d) and Figure 1 (h)).

# Appendix C

To create a simulation cell with an edge dislocation and periodic boundary conditions along the line direction (Y axis) and the Burgers' vector (X direction), the following steps were used:

1. A cell is created with double the intended height (Z-axis or the slip plane normal) i.e. X = 540 Å (with limits -270 to 270), Y = 100 Å (with limits -50 to 50) and Z = 540 Å (with limits -270 to 270) is created.
2. Two edge dislocations, spanning across the Y-axis and located midway along the X-axis are introduced at Z = -135 Å and Z = 135 Å respectively.

3. The cell is allowed to undergo potential energy minimisation with periodic boundary conditions applied along all directions.
4. The minimised cell is now cut into half along the Z direction i.e. only atoms in the region of Z > 0 Å are retained.
5. The cell with X = 540 Å (with limits -270 to 270), Y = 100 Å (with limits -50 to 50) and Z = 270 Å (with limits 0 to 270) is subject to potential energy minimisation again with periodic boundary conditions applied along the X and Y directions to obtain the simulation cell shown in Figure 2 (a).

# Appendix D

To determine the defect-dislocation interaction mechanism, a constant shear rate approach is used here where the block of immobile atoms at the top of cell are sheared at a constant rate while the block of the immobile atoms at the bottom of the cell are kept fixed (Figure 2 (a)). To examine the variation of pinning strength with applied shear rate, simulations with a range of shear rates were attempted for $He_2V$-SIA defect cluster of size 3 placed in the path of the gliding edge dislocation (Figure 2 (a)).

For the highest shear rate attempted of 0.003 $ps^{-1}$, there was an unphysical rise in temperature within the cell, causing the atomic positions to become disorganised. Shear rate of 0.0005 $ps^{-1}$ was also found to not show any distinguishable difference in the recorded shear stress, between the cases for with and without the defect in the dislocation path (with a peak stress of 420 MPa being reached in either case).

The pinning effect (i.e. a distinct difference in shear stress between the cases of with and without defect in the dislocation path) was first observed for a shear rate of 0.0001 $ps^{-1}$. Subsequently simulations at lower shear rates were attempted and the pinning strength was found to reduce with lowering shear rate before tending towards saturation for rates lower than $10^{-6}$ ps$^{-1}$ (Figure 6 (a)). A polynomial fitting approach was used to fit the data points observed for the different shear rates (fitting line and corresponding equation in Figure 6 (a)). The equation determined from curve fitting was then used to estimate the pinning strength at a much lower shear rate of $10^{-16}$ ps$^{-1}$ (i.e. $10^{-4}$ s$^{-1}$ which corresponds to the strain rate used in the nano-indentation test in the prior study [14]). A small difference of ~2% was noticed between the pinning strengths at rates $10^{-6}$ ps$^{-1}$ and $10^{-16}$ ps$^{-1}$.

This indicates that values estimated from using a shear rate of $10^{-6}$ ps$^{-1}$ have relatively little dependence on the shear rate. To further confirm this, simulations with He$_2$V-SIA defect clusters of sizes 4 and 10, obstructing the gliding dislocation, were also attempted at rates $10^{-4}$ ps$^{-1}$, $10^{-5}$ ps$^{-1}$, $10^{-6}$ ps$^{-1}$ and $10^{-7}$ ps$^{-1}$ (Figure 6 (c),(d)). It is seen that, similar to the observation for cluster size 3, for cluster sizes 4 and 10 also, the pinning strength tends towards saturation below shear rate $10^{-6}$ ps$^{-1}$. Based on the polynomial curve fitting obtained for the cases of cluster size 4 and 10, the pinning strength at rate $10^{-6}$ ps$^{-1}$, was found to be within 2% of the interpolated value at rate $10^{-16}$ ps$^{-1}$. Based on these observations, all simulations in this study were performed at a constant shear rate of $10^{-6}$ ps$^{-1}$, to achieve an optimum balance between accuracy of results and computation time.

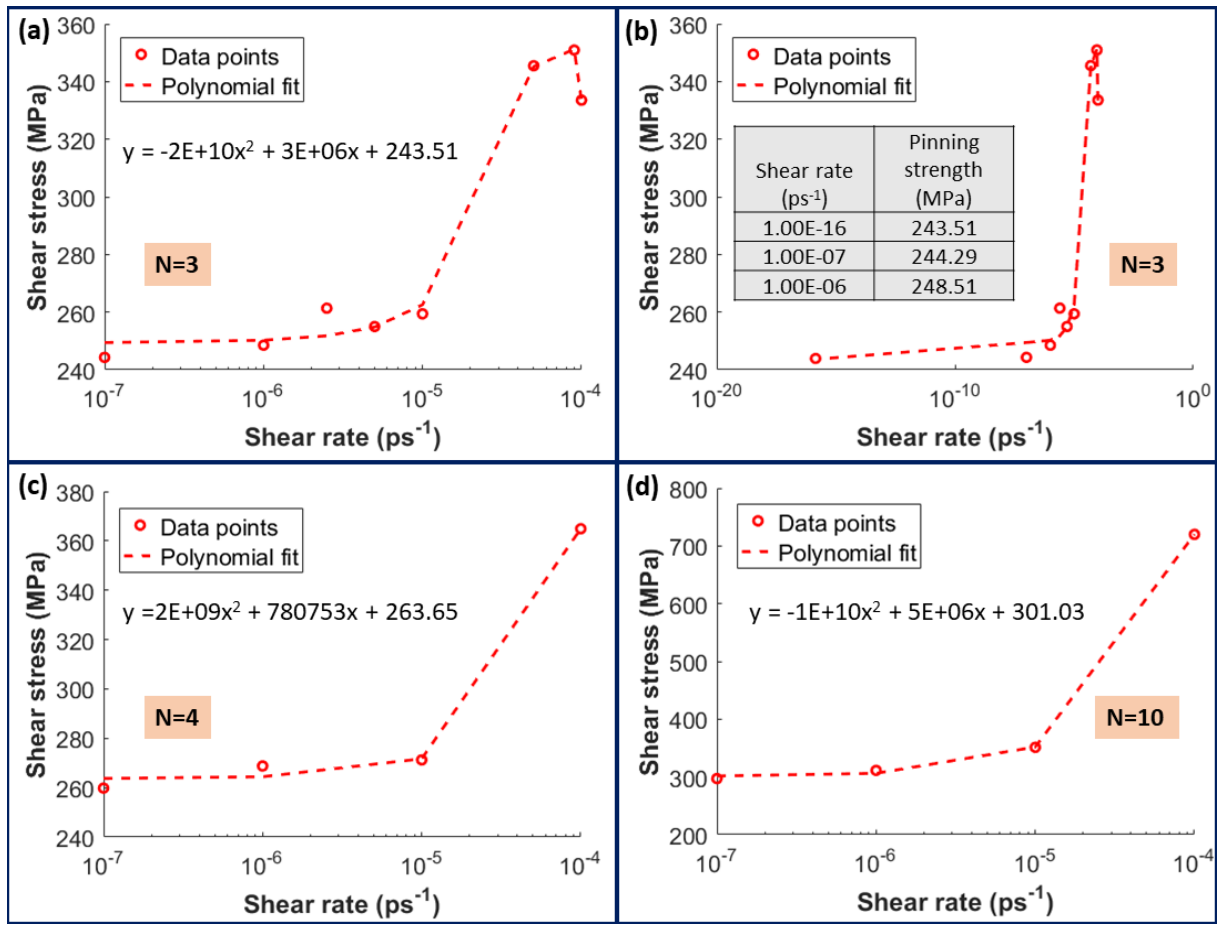

Figure 6 – Change of the pinning strength for $He_2V$-SIA defect cluster of size 3, with the applied shear rate (a) data points observed at the examined shear rates superimposed by the polynomial fit with the equation stated in the inset and (b) subplot (a) shown with the polynomial fit interpolated to estimate the pinning strength for shear rate $10^{-16}$ $ps^{-1}$. Pinning strength observed at the examined shear rates superimposed by the polynomial fit with the equation stated in the inset for (c) $He_2V$-SIA defect cluster of size 4 and (d) $He_2V$-SIA defect cluster of size 10.

# Appendix E

The pinning strength, i.e. the difference between the pinning force recorded for the defect-free case (averaged over the zone demarcated by the orange dotted line in Figure 2 (b)) and the maximum pinning force recorded with the defect present, varies as a function of the time over which the curves are averaged (Figure 7). Figure 7 shows that when the data is averaged over a period of 50 ps or more, the variation in the pinning strength diminishes and tends towards a saturated value of ~ 0.36 µN. Thus, an averaging time period of 100 ps is chosen and the uncertainty in the calculated pinning strength is measured as ±1 standard deviation across the five values estimated by considering five different averaging time intervals, i.e. 20, 50, 100, 150 and 250 ps (Figure 3 (k) and Table 1).

## Appendix F

The dimensions of the simulation cell are X = 540 Å (with limits -270 to 270), Y = 100 Å (with limits -50 to 50) and Z = 270 Å (with limits 0 to 270). The cell contains 944463 i.e. approximately a million atoms. Since the concentration of implanted helium is known to be 3000 appm, the simulation cell may be estimated to contain ~3000 helium ions.

Considering uniform distribution of defects within the cell implies uniform spacing between defects. Let the spacing between defects be $x$.

Thus, number of defects along X, Y and Z directions is $(540/x)$, $(100/x)$ and $(270/x)$ respectively.

For given cluster size N, the concentration of defect clusters will be $B = 3000/A$ appm where $A$ is the total number of helium ions in the cluster (Table 1).

Thus, total number of such clusters expected in the considered simulation cell is ~$B$. Therefore,

$$(540/x) \times (100/x) \times (270/x) = B \qquad (6)$$

Eq. (7) is used to compute the spacing between defects i.e. $x$ for a given cluster concentration $B$ as stated in the form of parameter *C* in Table 1.

The dislocation lies on the XY plane (Figure 2 (a)) and moves in the X direction. Thus, for a given cluster concentration $B$, at any point, the maximum number of defect clusters it may encounter will be $(100/x)$ i.e. the ratio of the magnitude of the cell in the Y direction to the spacing between defects (stated as parameter *D* in Table 1).